\documentclass[preprint,12pt]{elsarticle}

\begin{document}

\begin{frontmatter}
\title{Stable marriage problem under Monte Carlo simulations, influence of preferrence corelation on relaxation time}

\author{P. Nyczka, J. Cis{\l}o}

\address{Institute of Theoretical Physics, University of Wroc{\l}aw, pl. Maxa
Borna 9, 50-204 Wroc{\l}aw, Poland\\}


\begin{abstract}
In this paper we consider stable marriage problem under the Monte Carlo simulations. We investigate how correlation in lists of preferrences can affect simulation results such as: relaxation time, time distribution of relaxation times etc. We took into account attractiveness of individuals and it's different types as well as personal taste.
\end{abstract}


\end{frontmatter}

\section{Introduction}
\label{sec:introduction}

We revisited problem well known from the game theory: the stable marriage problem   \cite{bib:sex-oriented-stable-matchings}, which is also widely used in economics etc. In this problem two sets of agents (e.g. men and women) must be mathed pairwise in accordance to their mutual preferrences. Those preferrences could be in conflict, and agents are egoistic in addition. That means, each of them tries to maximise its own satisfaction (find the best partner) without respecting the rest. However there is virtually impossible to make all of the agents absolutely happy, there are some states where they are more less satisfied with their partners and can't change them anymore. These states are known as stable states or Nash equilibria. Precisely this equilibrium it is the situation, where there is no such two agents from opposite sets, whose both prefer to be together instead of staying with their actual partners in other words there is no unstable pairs. Usually there is several possible stable states in one set.

Most of simulations of matching problem use deterministic algorithms to find the stable states. It works well quick and elegant, but we wanted to focus on another aspect of stable marriage problem. We needed more realistic model, because we wanted to investigate system's relaxation time (not only find stable state itself) in "real life" situation, which is the time of reaching the stable state. We also wanted to know how this time can be affected by system size and correlations of the preferrences lists.

In deterministic "classical" case each agent from one set knows all the agents from opposite set, and have a list of them in order of preferrence. Algorithm chooses optimal order of encounters, to minimize time needed to reach the stable state. In real life, people don't know all of their potential partners at the beginning, and we can assume that people meet each other randomly. To simulate this we used Monte Carlo simulations. In this case the encounters between agents are random.

In real populations attractiveness of individuals differs from one to another. It's very important fact and we considered this, by introducing some kind of beauty factors which caused a correlations between the lists of preferrences. Our assumption was similiar to the one prestented in \cite{bib:beauty-and-distance-stable-matchings}, but we introduced beauty in different way. We noticed that people can have also different tastes and types of beauty, and simulated this by introducing attraciveness, and taste vectors, which seems to be more realistic and more detailed than in above paper.

In our survey we compared different correlations strength (different dimensions of attraciveness and taste vectors - the greater dimension, the weaker correlation) and uncorrelated case as well. Article is focused on influence of such correlations on the relaxaton time.

\section{Model}
\label{sec:model}

Described model is a simple model of stable marriage problem, mentioned above. The goal is to rich the stable state, but, as it has been said, we used Monte Carlo simulations instead of common used deterministic algorithm.

There are sets of $N$ men $M = \{m_1,m_2,m_3, ... ,m_N\}$ and $N$ women $W = \{w_1,w_2,w_3, ... ,w_N\}$. Each agent has it's own preference list of opposite sex representants. This is simply a subjective ranking list of attractivness of potential partners from opposite sex. Let's define two matrices: $P_m$ for men and $P_w$ for women, where $P_m(m_i,w_j)$ element denotes the rank of the woman $w_j$ on the ranking list of the man $m_i$, and the $P_w(w_j,m_i)$ denotes the rank of the man $m_i$ on the ranking list of the woman $w_j$. The lower rank, the higher position.

Agents from opposite sets must be matched pairwise to create collection of $R={(m,w)i}_{i=1,...,N}$ relatonships, to do this one have to arrange encounters between them. There are two different ways of this arrangement. In deterministic case - the order of encounters is determined by the preferrences lists, in MC case - order is random.

\subsection{Dynamics}
\label{subsec:dynamics}

In each time step we arrange random encounter between opposite sets representants.

During each encounter both of the potential partners check their mutual attractiveness and may declare a will to commit to a new partnership. An agent declares such will when is free (has no partner), or when potential new partner is higher on preferences list ($P$ is lower) than it's current partner . If both of them declare they willing, their break up their previous relationships (if they have ones) and the new one  between them is created.

Only if  $P_m(m_i',w_j') < P_m(m_i',w_j)$ or $m_i'$ has no partner,
and $P_w(w_j',w_i') < P_w(w_j',w_i)$ or $m_j'$ has no partner, the new relationship is created.
Where primmed are the people who ere doing the meeting, and unprimmed are their actual partners ($m_i$ is in relation with $w_i$ ${m_i,w_i}$ and $m_j$ with $w_j$ ${m_j,w_j}$).

The simulation ends while there is no such pair of agents, which can change it's partners. This state is called a stable state, or the Nash equilibrium. Usually there are several possible stable states for one set.

\subsection{Preferrences lists}
\label{subsubsec:preferrences}

We have made simulations for two common types of preferrences lists. First was the most classical case: random lists, and the second more realistic: correlated ones.

\subsubsection{Random lists}
\label{subsubsec:random-lists}

Random list construction is very simple. Each agent have such a list, with oppsite sex representants in random order. However it's simple, seems to be quite unrealistic \cite{bib:beauty-and-distance-stable-matchings}. As we know people differs in their attractiveness from person to person. Random lists don't take this fact into account.

\subsubsection{Correlated lists}
\label{subsubsec:correlated-lists}

Construction of correlated lists is little more complicated than in random case. People are different, their have different attacitveness and tastes as well. However that was claimed in \cite{bib:beauty-and-distance-stable-matchings}, we did it in somehow different way. Our approach takes into account different tastes, and different types of beauty. We can describe attracitveness by one number, but we can use several numbers (vector) to describe different aspects of its. Let $A(a_1,a_2,a_3, ... ,a_n)$  be a vector describing agents's attractiveness, and  $T(t_1,t_2,t_3, ... ,t_n)$ will be another vector to describe individual's taste, where $n$ is number of those aspects. Numbers in $A$ are real and randomly chosen from division $<0,1>$, and could describe different qualities of attractiveness, such as beauty, intelligence etc. Numbers in $T$ are describing attention paid to given quality in potential mate evaluation, or in other words: "weigths" and they are different for different agents. On the beginning they are also randomly chosen real numbers from division $<0,1>$, but this vector is to be normalised later, so $|T| = 1$.  

When there are $A$ and $T$ vectors, created for all the agents, we have to make preferrences lists. Whole procedure is quite simple. An agent $i$ evaluates all agents from opposite set one by one and sort them by obtained score $S$, the higher score - the higher place on preferrences list of an agent. Where evaluation of agent $j$ by agent $i$ is just a simple scalar product: $S = T_i \cdotp A_j$. When preferrences lists are ready, simulation starts, and runs until stable state is reached.

\section{Results}
\label{sec:results}

\begin{figure} [!ht]
\begin{center}
\includegraphics{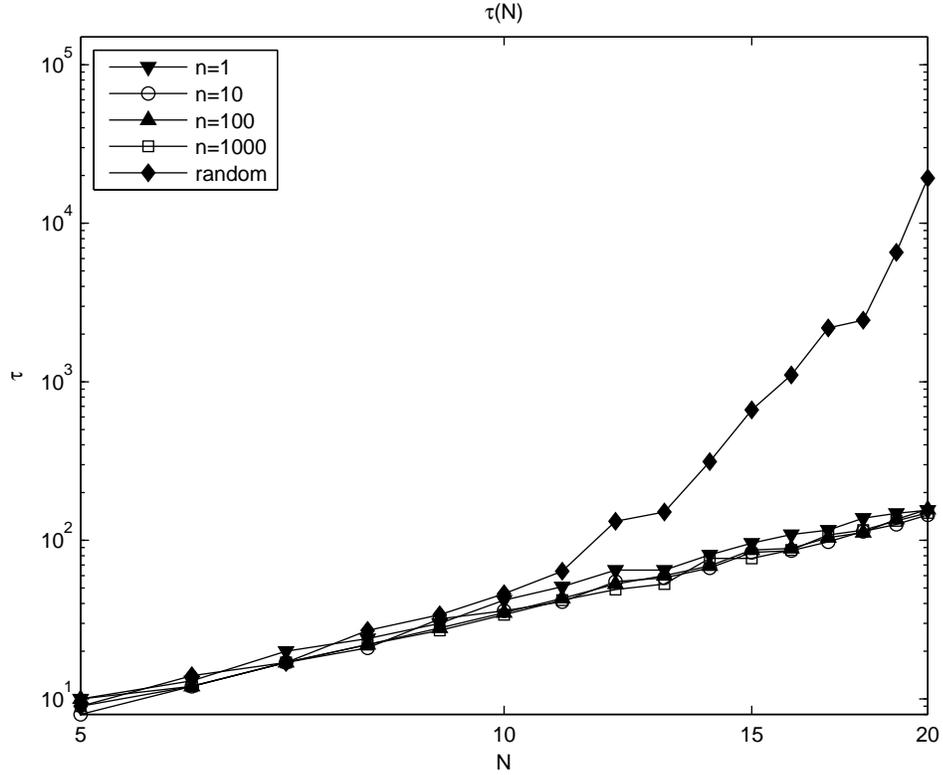}
\caption{Median of $\tau(N)$ for different types of preferrences lists in log-log scale, where $n$ is number of qualities decribing attractivness. Each result was is average of 100 MC simulations.}
\label{fig:tau_n}
\end{center}
\end{figure}

We investigated relaxation times $\tau$, which means number of MC steps needed to reach a stable state. It was done for different number of agents, and for different types of preferences lists. In general there was two common types of that lists: correlated and random ones. As we found, $\tau$ strongly depends on list type (see fig.~\ref{fig:tau_n}).

For strongly correlated preferences lists $\tau$ is very small . Extremal case of correlation is situation where agents from one sex have identical lists. It takes place when $n=1$. For random lists, agents have different lists and $\tau$ is much greater.

In case of correlated lists when $n$ is greater, also the $\tau$ is greater, but for higher $n$ differences are smaller. Even for $n>1000$ there is still huge gap between $\tau$ for correlated and random lists. In correlated case function $\tau(N)$ satisfies power law for some $N < N_c$ and then grows much faster.  For $n=1$ power law is satisfied for all(?) $N$. Suprising is fact that for some system's size there is optimal $n = n_o$ value for which relaxation time is minimal, for $N > 350, n_o = 1$, but for $50 < N < 350, n_o = 2$. 

Suprising is fact that even for such small $N$ as $N=20$ the size of $\tau$ gap between uncorellated and correlated cases is about $10^3$ ,which is really spectacular difference.

\begin{figure} [!ht]
\begin{center}
\includegraphics{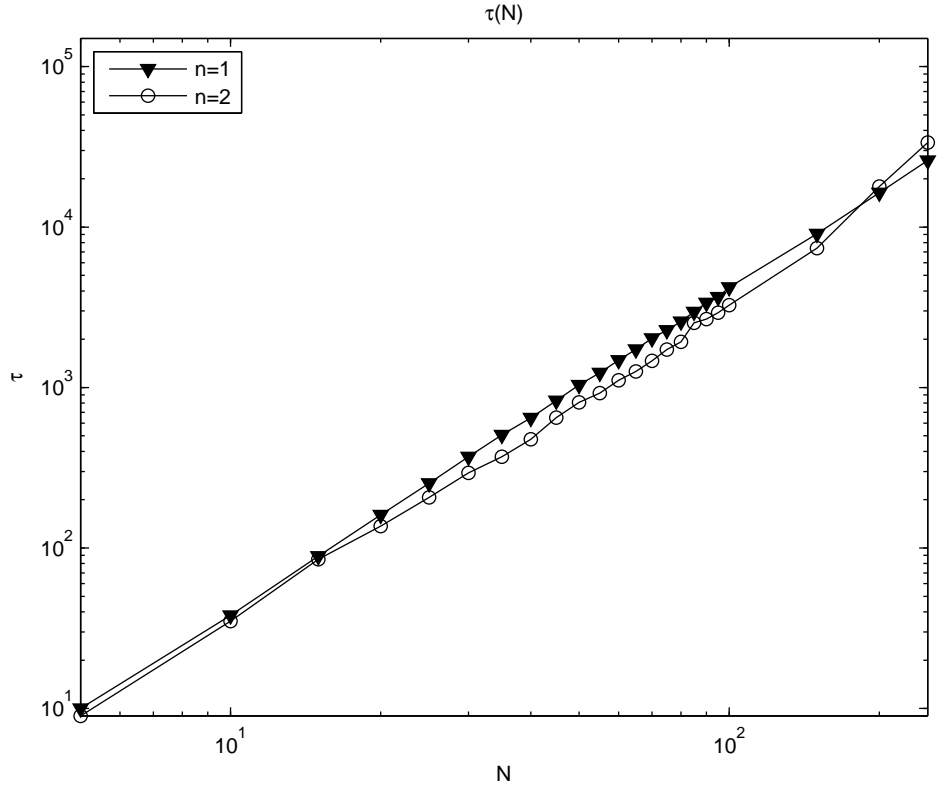}
\caption{Median of $\tau(N)$ for two different $n$ values log-log scale, where $n$ is number of qualities decribing attractivness. Each result was is average of 100 MC simulations.}
\label{fig:tau_n_1_2}
\end{center}
\end{figure}

\section{Discussion}
\label{sec:discussion}

As it has been shown, the more correlated lists the shorter relaxation times $\tau$ for $N > 350$.

In other words: the more common tastes or more different individuals are, the shorter times occur. One can claim a hypothesis that for shorter mating times, also reproduction is more effective. An assumption that only stable marriages can have children, implies that younger couple can have more children thand older ones. Therefore natural selection could prefer some kind of common taste and diversity of individuals, to maximise offspring count and minimise cost of sexual selection.

For smaller $N$ there are some other differences. Suprising is fact that for smaller groups this time is minimal for $n > 1$   (see fig.~\ref{fig:tau_n_1_2}) which means weaker correlation, stronger personalisation. For some reason in small groups, some weakening of correlation which means more idividualisation, causes quicker mathing.

\end{document}